\documentclass[aps,pra,amsmath,amssymb,letterpaper,twocolumn]{revtex4}

\usepackage{hyperref}
\usepackage{graphicx}
\usepackage{color}

\begin{document}

\title{Approaching the Full Configuration Interaction Ground State
  from an Arbitrary Wavefunction with Gradient Descent and
  Quasi-Newton Algorithms}

\author{Carlos A. Jim\'enez-Hoyos}
\email{cjimenezhoyo@wesleyan.edu}
\affiliation{Department of Chemistry, Wesleyan University, Middletown,
  CT, 06459}

\date{\today}

\begin{abstract}
We consider gradient descent and quasi-Newton algorithms to optimize
the full configuration interaction (FCI) ground state wavefunction
starting from an arbitrary reference state $|0 \rangle$. We show that
the energies obtained along the optimization path can be evaluated in
terms of expectation values of $|0 \rangle$, thus avoiding explicit
storage of intermediate wavefunctions. This allows us to find the
energies after the first few steps of the FCI algorithm for systems
much larger than what standard, deterministic FCI codes can handle at
present. We show an application of the algorithm with reference
wavefunctions constructed as linear combinations of non-orthogonal
determinants.
\end{abstract}

\maketitle

\section{Introduction}

In principle, all the chemical properties of molecular systems can be
determined from knowledge of the eigenstates of the Hamiltonian
operator. The full-configuration interaction (FCI) wavefunction
constitutes a prominent paradigm of quantum chemistry
\cite{szabo_ostlund}: it provides the exact solution of the
Schr\"odinger equation on a basis of suitably selected $N$-electron
wavefunctions. While finding the FCI ground state wavefunction is
algorithmically easy, the number of degrees of freedom in it increases
exponentially with system size. Therefore, one of the central goals of
the quantum chemistry community is to develop approximations that can
yield properties of similar quality as that from the FCI ground state
while reducing significantly the computational cost entailed.

The optimization of the FCI ground state wavefunction is usually cast
as an eigenvalue problem. Given a matrix representation of the
Hamiltonian operator $H$ and a vector representation of a starting
reference wavefunction $|0 \rangle$, potent algorithms (with that from
Davidson \cite{davidson1975} being a prominent example) have been
developed to optimize the FCI ground state wavefunction
$|\Psi_\mathrm{FCI} \rangle$ while minimizing the number of
matrix-vector operations and the number of intermediate
wavefunctions. Nevertheless, the dimension of the FCI problem renders
this approach possible only for relatively small systems.

In this work, we obtain the FCI ground state from an unconstrained
optimization of the wavefunction amplitudes in the orthogonal
complement of some reference state $|0 \rangle$. We carry the
unconstrained optimization with gradient descent and quasi-Newton
algorithms. In this way, the reference wavefunction is improved
systematically with increasing number of steps. The central result
obtained in this work is that the energies obtained along the
optimization path can be fully expressed in terms of matrix elements
of the reference $|0 \rangle$.  Explicitly casting the energy after
some finite number of steps as a functional of the reference
wavefunction allows us to consider systems for which the dimension of
the Hilbert space is much larger than the disk or memory available in
current computational facilities.

There is another important reason to avoid the explicit vector
representation of the FCI wavefunction, which actually constitutes the
motivation for this manuscript. Namely, it allows us to consider
classes of wavefunctions for which a vector representation in some
convenient orthonormal $N$-electron basis is not available, and the
computational cost of evaluating it would be proportional to the
dimension of the Hilbert space. A specific example that we shall
consider are reference states written as linear combination of a few
determinants that are, in general, non-orthogonal ({\em i.e.},
$\langle \Phi_1 | \Phi_2 \rangle \neq 0$):
\begin{equation}
  |0 \rangle = \sum_q f_q |\Phi_q \rangle \nonumber.
\end{equation}
In that case the orthogonal complement is not known {\em a priori} and
the vector representation in a basis of, {\em e.g.}, orthonormal
Slater determinants can only be determined at great computational
expense. This class of wavefunctions occur in non-orthogonal
configuration interaction (NOCI) \cite{sundstrom2014} as well as in
symmetry-projected Hartree--Fock methods \cite{jimenez2012}.

The rest of this manuscript is organized as follows. In
Sec. \ref{sec:theory} we describe how we parametrize the FCI ground
state wavefunction and describe the gradient descent and quasi-Newton
optimization algorithms considered. In Sec. \ref{sec:results} we
discuss the application of the method in an H$_4$ ring and in the
determination of spectroscopic constants of the N$_2$, O$_2$, and
F$_2$ molecules. Finally, in Sec. \ref{sec:conclusions} we provide
some closing remarks.

\section{Theory}
\label{sec:theory}

We parametrize the FCI ground state $|\Psi_\mathrm{FCI} \rangle$ using
an exponential, non-Hermitian ansatz of the form
\begin{align}
  |\Psi \rangle &= \, \exp (\hat{Z}) |0 \rangle, \\
  \hat{Z} &= \, \sum_x Z_x |x \rangle \langle 0|,
\end{align}
where $|0 \rangle$ is the reference wavefunction and $|x \rangle$
labels an orthonormal state in the orthogonal complement of $|0
\rangle$. The FCI ground state can be reached after an unconstrained
optimization in the parameters $Z$. (We note that the same
parametrization was used in Ref. \onlinecite{wang2019}, also casting
FCI as an unconstrained minimization problem.) An important ingredient
of this work is that an explicit construction of the states $\{ |x
\rangle \}$ is not necessary. We assume in what follows that $\langle
0 | \Psi_\mathrm{FCI} \rangle \neq 0$.

Given $Z$ (as a vector of amplitudes), the energy of $|\Psi \rangle$
can be evaluated as
\begin{align}
  E[Z] &\equiv \, \frac{\langle \Psi | H | \Psi \rangle}{\langle \Psi |
    \Psi \rangle} = \frac{\langle 0| \exp(\hat{Z}^\dagger) H
    \exp(\hat{Z}) |0 \rangle}{\langle 0| \exp(\hat{Z}^\dagger)
    \exp(\hat{Z}) |0 \rangle} \nonumber \\
  &=\, \frac{H_0^0 + H_0^x Z_x + Z^x H_x^0 +
    Z^x H_x^y Z_y}{1 + Z^x Z_x},
\end{align}
where $H_\alpha^\beta = \langle \alpha | H | \beta \rangle$, Einstein
summation is implied and the indices $x,y$ run only over the
orthogonal complement of $|0 \rangle$. Here, we have used the fact
that $\hat{Z^2} |0 \rangle = 0$. We have also assumed a real
wavefunction $|0 \rangle$ and real coefficients $Z$, as we do
throughout this work.

The energy gradient with respect to the amplitudes in $Z$, evaluated
at $Z=Y$ is given by
\begin{equation}
  g_x \equiv \left. \frac{\partial E[Z]}{\partial Z^x}
  \right|_{Z=Y} = \frac{2H_x^0 + 2H_x^y Y_y - 2E[Y] Y_x}{1 + Y^y Y_y}.
  \label{eq:grad}
\end{equation}

Along the optimization paths described in the next subsections a line
search is needed to minimize the energy of $E[Y+sX]$, with $Y$ being
the vector of current amplitudes and $X$ the search direction, with
respect to the step size $s$. Explicitly, $E[Y+sX]$ becomes
\begin{widetext}
\begin{equation}
  E[Y+sX] = \frac{E[Y](1 + Y^x Y_x) + s (H_0^x + Y^y H_y^x) X_x + s
    X^x (H_x^0 + H_x^y Y_y) + s^2 X^x H_x^y X_y}{(1 + Y^x Y_x) + s Y^x
    X_x + s X^x Y_x + s^2 X^x X_x}.
\end{equation}
\end{widetext}
This is a rational equation in $s$ of the form
\begin{equation*}
  \frac{c + b s + a s^2}{f + e s + d s^2}.
\end{equation*}
Let $s_\ast$ be the value of $s$ that extremizes $E[Y+sX]$; this takes
the form
\begin{equation*}
  s_\ast = \frac{(cd-af) \pm \sqrt{(cd-af)^2 -
      (ae-bd)(bf-ce)}}{(ae-bd)}.
\end{equation*}

For convenience, we shall introduce the quantities $f_1 = H_0^0$, $f_2
= H_0^y H_y^0$, $f_3 = H_0^y H_y^z H_z^0$, etc. Note that all those
matrix elements can be evaluated in terms of expectation values from
the reference wavefunction $|0 \rangle$. For instance,
\begin{align}
  f_2 &= \, \langle 0| H^2 |0 \rangle - f_1 \langle 0| H |0 \rangle, \\
  f_3 &= \, \langle 0| H^3 |0 \rangle - f_1 \langle 0| H^2 |0 \rangle
  - f_2 \langle 0| H |0 \rangle.
\end{align}

\subsection{Gradient Descent}

We begin at $Z_0 = 0$ with $|\Psi_0 \rangle = |0\rangle$. The gradient
at $Z_0$ is
\begin{align}
  (g_0)_x &= \, \alpha_0 H_x^0,
\end{align}
with $\alpha_0 = 2$.

In a standard gradient descent implementation a line search would be
performed along $-g_0$. It is common practice to accept a step size
that satisfies Wolfe \cite{wolfe1969,wolfe1971} conditions to avoid
the potentially expensive full line search. In our case, we aim to
perform a full line search, as we may only be able to afford a few
steps. The optimal step size $\sigma_\ast$ can be easily found given
the rational form of $E[Z]$ discussed above. Therefore, $Z_1 =
-\sigma_\ast g_0$ and $|\Psi_1 \rangle$ becomes
\begin{equation}
  |\Psi_1 \rangle = e^{-\sigma_\ast g_0} |0 \rangle.
\end{equation}

An explicit expression for $E_1$ is given by
\begin{equation}
  E_1 \equiv \frac{\langle \Psi_1 | H | \Psi_1 \rangle}{\langle
    \Psi_1 | \Psi_1 \rangle}
  = \frac{f_1 - 2 \sigma_\ast \alpha_0 f_2 + \sigma_\ast^2 \alpha_0^2 f_3}{1 +
    \sigma_\ast^2 \alpha_0^2 f_2},
  \label{eq:e1}
\end{equation}
with
\begin{equation}
  \sigma_\ast = \frac{(4f_3 - 4f_1 f_2) \pm \sqrt{64f_2^3 + (4f_3 - 4f_1
      f_2)^2 }}{-16f_2^2}.
\end{equation}
Note that $E_1$ is a functional of $|0 \rangle$ which can be
determined after evaluation of $f_1$, $f_2$, and $f_3$. (This is an
interesting functional in itself. One may consider optimizing $E_1$
with respect to the reference wavefunction parameters $|0 \rangle$,
which we have done for one of the examples considered.)

We now proceed to take another step. The gradient at $Z_1 =
-\sigma_\ast g_0$ is given by
\begin{equation}
  (g_1)_x =  \alpha_1 H_x^0 + \beta_1 H_x^y H_y^0
  \label{eq:g1}
\end{equation}
with
\begin{align}
  \alpha_1 &= \, \frac{2 (1 + \sigma_\ast \alpha_0 E_1)} {1+\sigma_\ast^2 \alpha_0^2 f_2}, \\
  \beta_1 &= \, \frac{-2\sigma_\ast \alpha_0}{1+\sigma_\ast^2 \alpha_0^2 f_2}.
\end{align}

After looking for the optimal step size $\tau_\ast$ (i.e., minimizing
$E[-\sigma_\ast g_0 - \tau g_1]$ with respect to $\tau$), $|\Psi_2
\rangle$ becomes
\begin{equation}
  |\Psi_2 \rangle = e^{-\sigma_\ast g_0 - \tau_\ast g_1} |0 \rangle.
\end{equation}

An explicit expression for $E_2$ is given below
\begin{widetext}
\begin{equation}
  E_2 \equiv \frac{\langle \Psi_2 | H |
  \Psi_2 \rangle}{\langle \Psi_2 | \Psi_2 \rangle}
  = \, \frac{E_1 (1 + \sigma_\ast^2 \alpha_0^2 f_2) - 2\tau_\ast (\alpha_1 f_2
    + \beta_1 f_3) + 2 \sigma_\ast \tau_\ast \alpha_0 (\alpha_1 f_3 + \beta_1
    f_4) + \tau_\ast^2 (\alpha_1^2 f_3 + 2\alpha_1 \beta_1 f_4 +
    \beta_1^2 f_5)}{(1 + \sigma_\ast^2 \alpha_0^2 f_2) + 2 \sigma_\ast
    \tau_\ast \alpha_0 (\alpha_1 f_2 + \beta_1 f_3) + \tau_\ast^2 (\alpha_1^2
    f_2 + 2\alpha_1 \beta_1 f_3 + \beta_1^2 f_4)}.
  \label{eq:e2}
\end{equation}
\end{widetext}
A closed-form expression for $\tau_\ast$ can be deduced from the
rational form of $E_2$ as a function of $\tau$, as described
before. $E_2$ is also a functional of $|0 \rangle$ that can be
assembled after evaluation of $f_1$, \ldots, $f_5$.

A third step would require the evaluation of $g_2$. Given the form of
the gradient (Eq. \ref{eq:grad}), it can be shown that $g_2$ takes the
form
\begin{equation}
  (g_2)_x = \alpha_2 H_x^0 + \beta_2 H_x^y H_y^0 + \gamma_2 H_x^y
  H_y^z H_z^0.
\end{equation}
Therefore, $E_3$ would be also a functional of $|0 \rangle$ that can
be determined after evaluation of $f_1$, \ldots, $f_7$. Subsequent
steps require the evaluation of higher order $f_k$ values.

At this point we note that the gradient descent approach described
above can, in principle, be used to converge to the true eigenfunction
of the Hamiltonian $\mathcal{H}$ as opposed to its representation in
some finite $N$-electron basis $H$. It would involve replacing the
expectation values $\langle 0| H^k |0 \rangle$ with $\langle 0|
\mathcal{H}^k |0 \rangle$, with everything else holding. This,
however, requires further work given that individual integrals
appearing in $\mathcal{H}^2$ (and higher powers) diverge when
evaluated with atomic Gaussian basis functions (see, {\em e.g.},
Ref. \onlinecite{nakatsuji2012}).

As a second remark, we note that the evaluation of other expectation
values can be done in the same way as the energy. For instance, the
expectation value of $W$, after the first gradient descent step, is
given by
\begin{equation}
  X_1 \equiv \frac{\langle \Psi_1 | W | \Psi_1 \rangle}{\langle \Psi_1
    | \Psi_1 \rangle} = \frac{f^w_1 - 2 \sigma_\ast \alpha_0 f^w_2 +
    \sigma_\ast^2 \alpha_0^2 f^w_3}{1 + \sigma_\ast^2 \alpha_0^2 f_2},
\end{equation}
with
$f^w_1 = W_0^0$, $f^w_2 = \frac{1}{2} (W_0^y H_y^0 + H_0^y W_y^0)$,
$f^w_3 = H_0^y W_y^z H_z^0$. This is also a functional of $|0 \rangle$
since
\begin{align}
  f^w_2 &= \, \textstyle \frac{1}{2} \langle 0| WH + HW |0 \rangle -
  f_1 \langle 0| W |0 \rangle, \\
  f^w_3 &= \, \langle 0| HWH |0 \rangle - \textstyle \frac{1}{2} f_1
  \langle 0| WH + HW |0 \rangle \nonumber \\
  &-\, f^w_2 \langle 0| H |0 \rangle.
\end{align}

\subsection{Quasi-Newton}

Because the convergence of the gradient descent algorithm is slow, we
now consider a quasi-Newton optimization. In a quasi-Newton approach
\cite{nocedal} it is customary to start from an initial inverse
Hessian $B_0$, which need only be an approximation to the true
Hessian. At each step along the optimization, the search direction
$p_k$ is determined from $B_k p_k = -g_k$, rather than setting $p_k =
-g_k$ as in the gradient descent approach. In this work we choose to
set $B_0 = I$ to keep things as simple as possible: under such
conditions, a quasi-Newton algorithm can still be fully defined in
terms of $|0 \rangle$. Naturally, if a better $B_0$ is used the number
of iterations required to reach convergence is expected to decrease.

Just as in the gradient descent approach, after defining a search
direction $p_k$ a line search is performed along it. Commonly, a step
size that satisfies Wolfe conditions is accepted. We choose, however,
to carry a full line search as in the gradient descent algorithm.

Because $B_0 = I$, it follows that the first step coincides with that
from the gradient descent approach, and $E_1$ remains unchanged. The
gradient $g_1$ is also the same as in gradient descent (see
Eq. \ref{eq:g1}). The direction $p_1$ is determined from $p_1 = -B_1
g_1$, with $B_1$ constructed using a quasi-Newton update formula.

We show in appendix \ref{sec:bfgs} that $p_1$, determined from a
Broyden-Fletcher-Goldfarb-Shanno (BFGS)
\cite{broyden,fletcher,goldfarb,shanno} update formula, takes the form
\begin{equation}
  (p_1)_x = \alpha'_1 H_x^0 + \beta'_1 H_x^y H_y^0,
\end{equation}
with $\alpha'_1$ and $\beta'_1$ being some constants that are
numerically different than $\alpha_1$ and $\beta_1$. Given that $p_1$
takes the same functional form as $g_1$, we conclude that $E_2$
determined from the BFGS approach is also a functional of $f_1$,
\ldots, $f_5$. Namely, $E_2$ would take the same form as
Eq. \ref{eq:e2}, with $\alpha_1 \to \alpha'_1$ and $\beta_1 \to
\beta'_1$. Further quasi-Newton steps can also be cast as functionals
of $|0 \rangle$, in the same way as in the gradient descent algorithm.

\section{Results and Discussion}
\label{sec:results}

We proceed to discuss the application of the optimization algorithms
described above in a H$_4$ ring and in the N$_2$, O$_2$, and F$_2$
diatomics.

\subsection{H$_4$ ring}

\begin{figure}[!htb]
  \includegraphics[width=6cm]{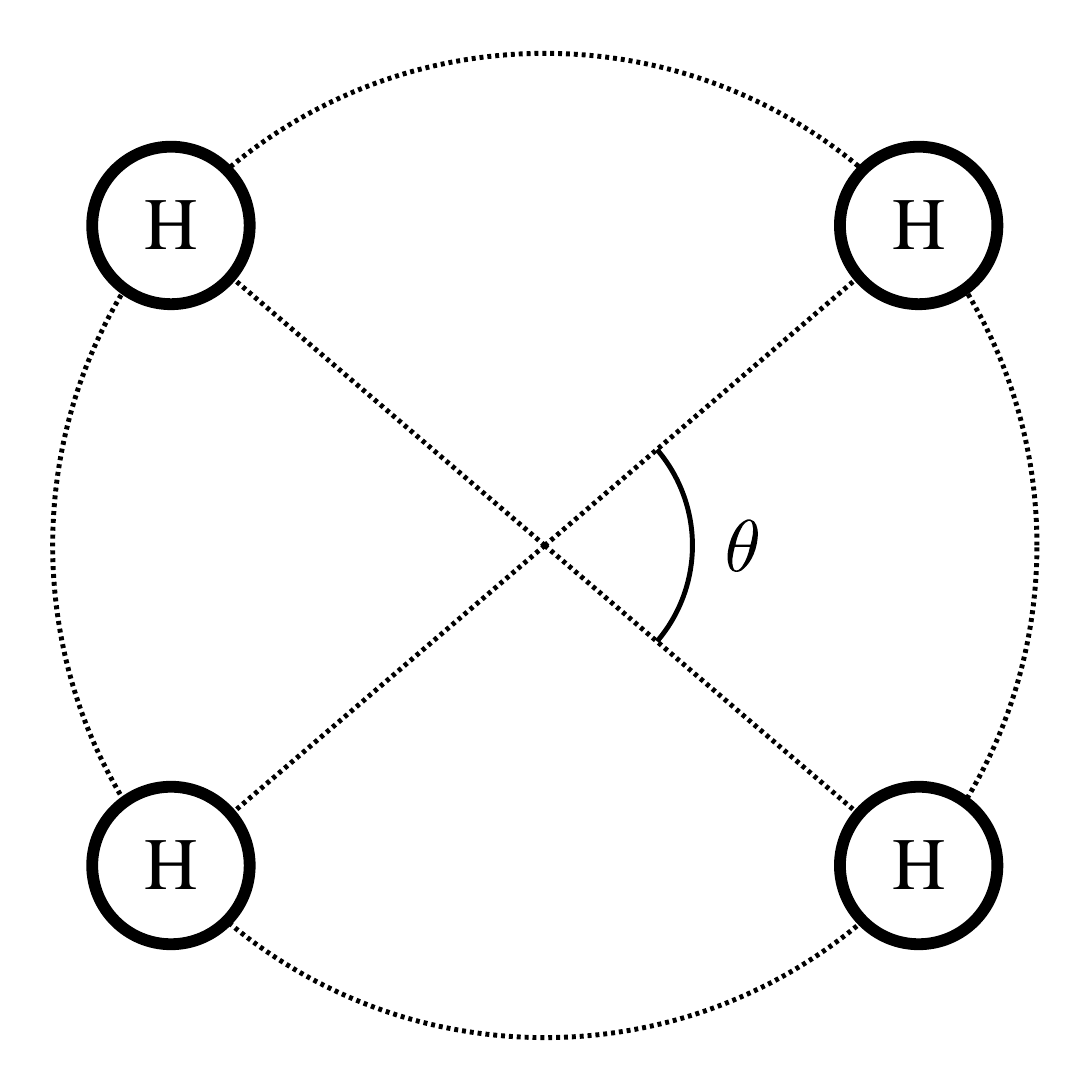}
  \caption{The H$_4$ system described in
    Ref. \onlinecite{scuseria2017} consists of four H atoms placed
    along a ring of radius $r=3.3$~bohr, controlled by an angle
    $\theta$. At $\theta \approx 24$~deg the system corresponds to two
    weakly interacting H$_2$ units near their equilibrium
    geometry. \label{fig:h4s}}
\end{figure}

We consider a system of 4 H atoms placed along a ring of radius
$r=3.3$~bohr \cite{scuseria2017}, with the arrangement depicted in
Fig. \ref{fig:h4s}. For $\theta \approx 24$~deg, the system consists
of two weakly interacting H$_2$ molecules near their equilibrium
geometry. Conversely, for $\theta = 90$~deg the 4 H atoms form a
square and the system has a strong multireference character. In
scanning $\theta$ in the range between 20~deg and 90~deg the system
evolves from a weak to a strong correlation regime. Our calculations
in this system were performed by explicitly constructing the vector
representation of the considered wavefunctions; this allows us to
study the convergence behavior of the gradient descent and
quasi-Newton algorithms.

\begin{figure*}[!htb]
  \includegraphics[width=8cm]{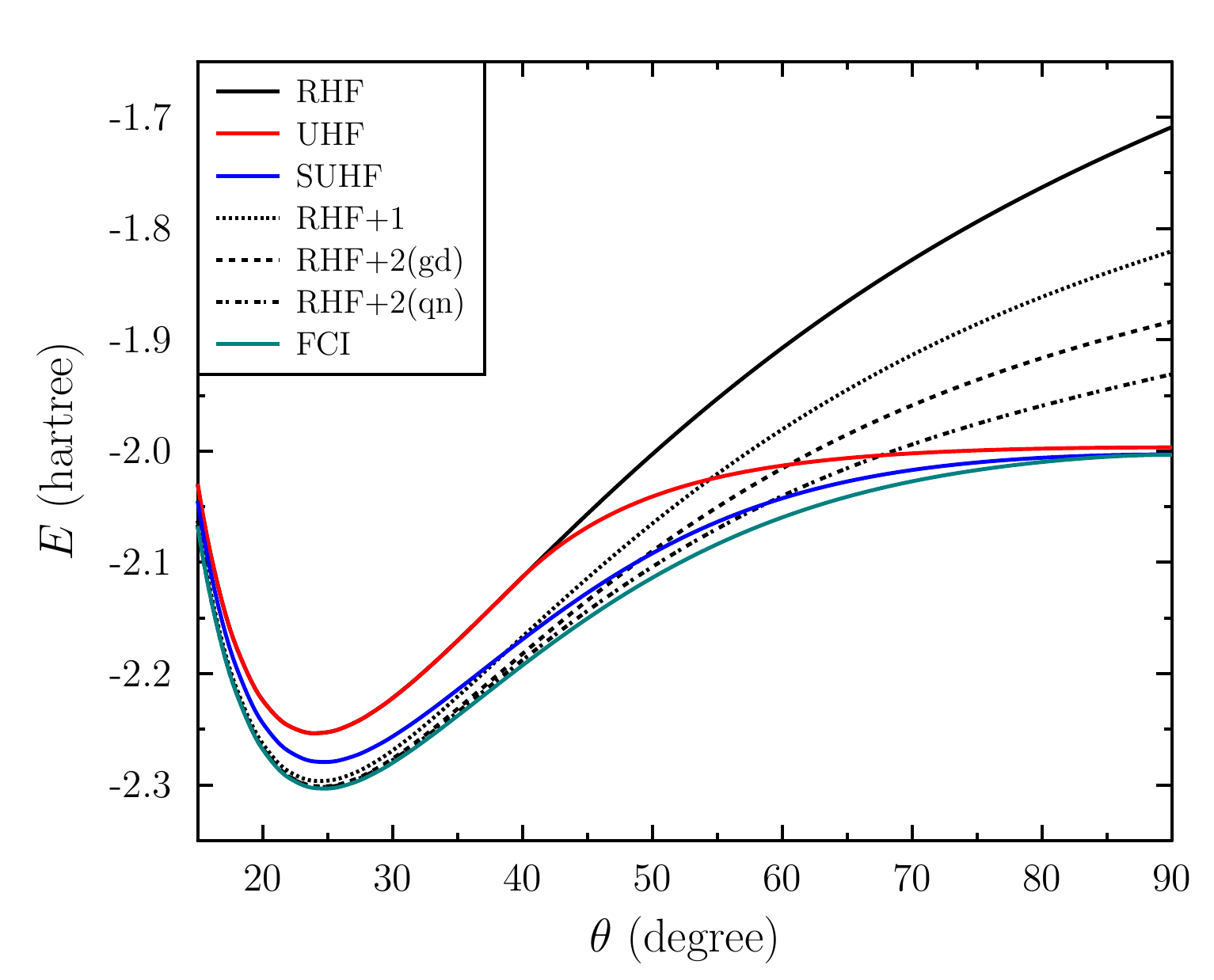}
  \hspace{0.2cm}
  \includegraphics[width=8cm]{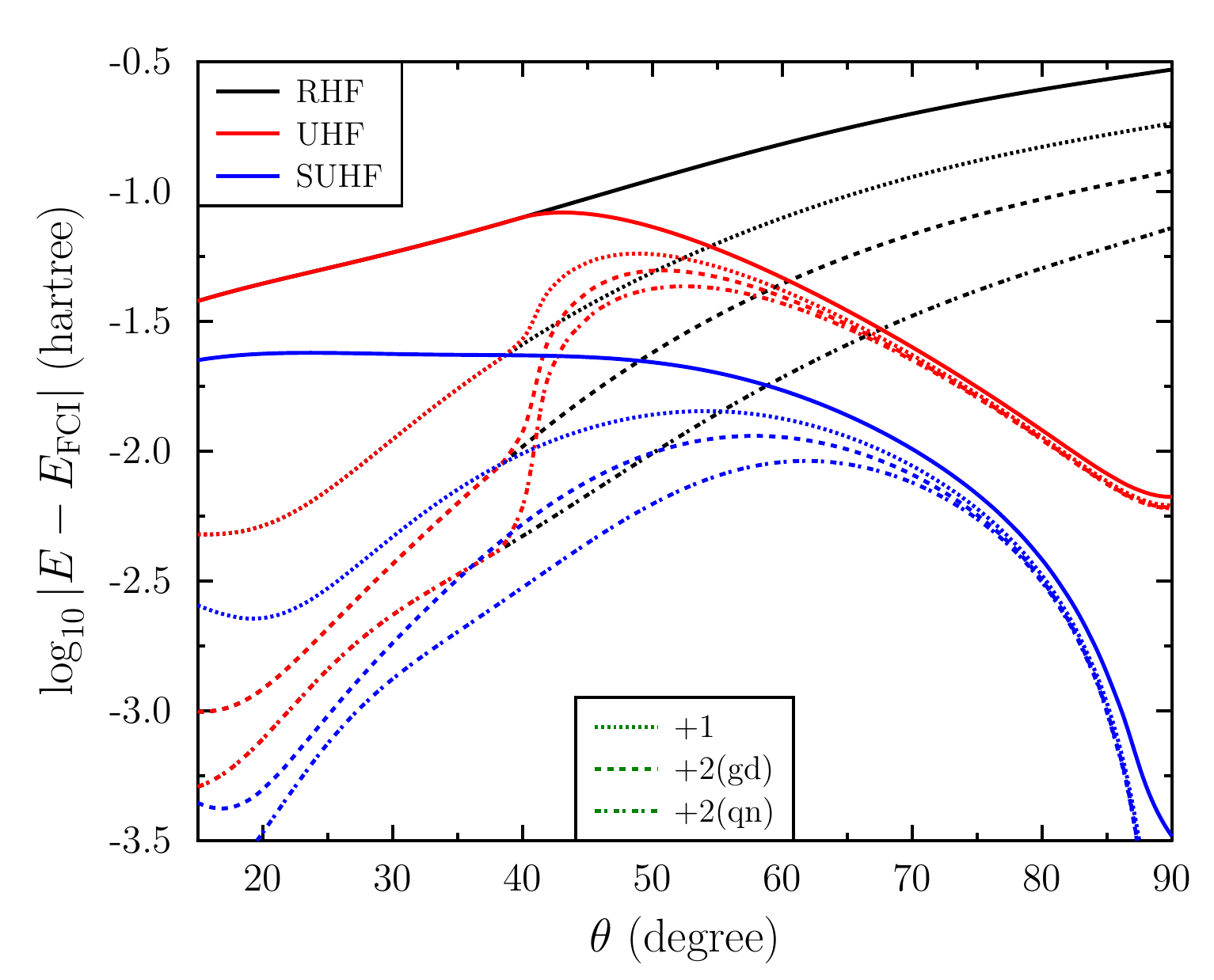}
  \caption{(Left) Energy (in hartree) of the H$_4$ system of
    Fig. \ref{fig:h4s}, as a function of $\theta$, computed with
    various methods and the 6-31G basis set. We show the energies
    after 1 and 2 gradient descent (gd) steps, as well as the energy
    after 2 quasi-Newton (qn) steps. (Right) Error in the energy, with
    respect to the FCI ground state, after 1 and 2 gd or qn steps
    starting from RHF, UHF, and SUHF wavefunctions. \label{fig:h4}}
\end{figure*}

We show in Fig. \ref{fig:h4} (left) the energy of the H$_4$ system as
a function of $\theta$, evaluated with the 6-31G basis set. In
particular, we show restricted and unrestricted Hartree--Fock (RHF and UHF),
spin-projected UHF (SUHF), and FCI ground state energies. In addition,
we show the energy after the first and second gradient descent (gd)
steps as well as the energy after the second quasi-Newton step
starting from the RHF wavefunction. On the right we show the errors
with respect to the FCI ground state calculated with each method,
where we now show also the corrections starting from UHF and SUHF. The
unusual profile displayed by the UHF curves is associated with the
collapse back to the RHF wavefunction at $\theta \approx 40$~deg.

The improvement upon the RHF energy obtained with the first gd step is
quite significant: the error is reduced by nearly a third at $\theta =
90$~deg and by an order of magnitude in the equilibrium region. The
second step (either gd or qn) reduces the energy even further, with
the qn step being considerably better than the gd step. In the case of
UHF and SUHF, the improvement after the first and second steps is very
modest near $\theta = 90$~deg, although the energy lowering near the
equilibrium region is still substantial. Around $\theta \approx
50$~deg, the first step correction to RHF leads to a lower energy than
the first step correction to UHF even when the reference wavefunction
itself is lower in energy. As more steps are taken, the expectation is
that the qn algorithm will outperform the gd algorithm as the former
converges linearly while the latter should approach quadratic
convergence \cite{nocedal}.

\begin{figure}[!htb]
  \includegraphics[width=8.5cm]{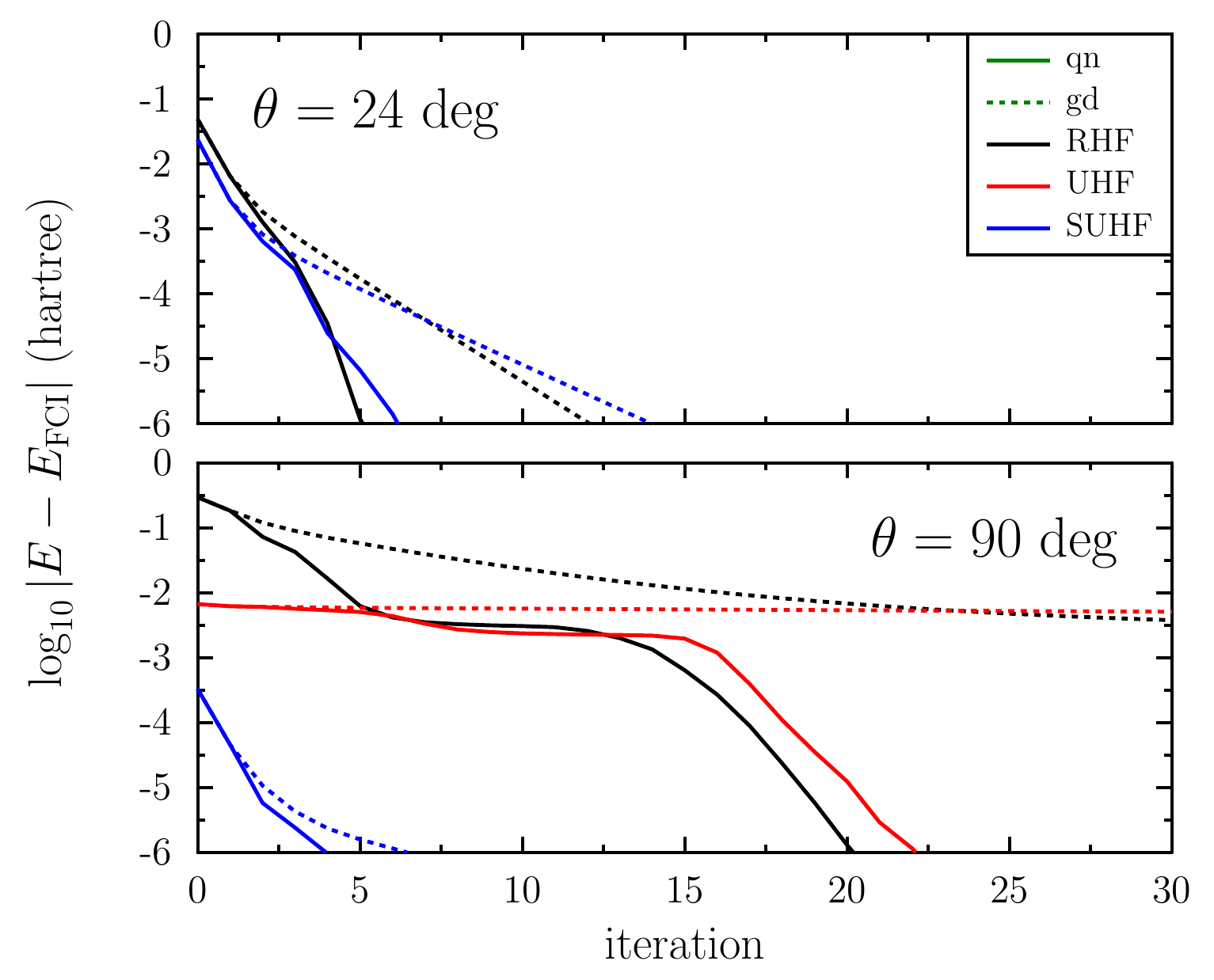}
  \caption{Convergence of the energy, at $\theta = 24$~deg (top) and
    $\theta = 90$~deg (bottom), as a function of iteration, using the
    gradient descent (gd) and quasi-Newton (qn) algorithms starting
    from the RHF, UHF, and SUHF reference
    wavefunctions. \label{fig:conv}}
\end{figure}

We show in Fig. \ref{fig:conv} the convergence profile of the gd and
qn algorithms on top of the RHF, UHF, and SUHF wavefunctions at
$\theta = 24$~deg (top) and $\theta = 90$~deg (bottom). At $\theta =
24$~deg, convergence is relatively fast: with gd, 10 steps are
sufficient to converge the energy to $10^{-5}$~hartree, while 5 steps
are enough with the qn algorithm. At $\theta = 90$~deg the profiles
are very different. While convergence starting from the SUHF reference
wavefunction is quite fast, a starting RHF or UHF wavefunction leads
to much slower convergence. In particular, the gd algorithm from UHF
converges extremely slowly. The qn algorithms do recover a faster
convergence rate after 15 or so iterations.

\subsection{N$_2$, O$_2$, and F$_2$}

\begin{figure}[!htb]
  \includegraphics[width=8.5cm]{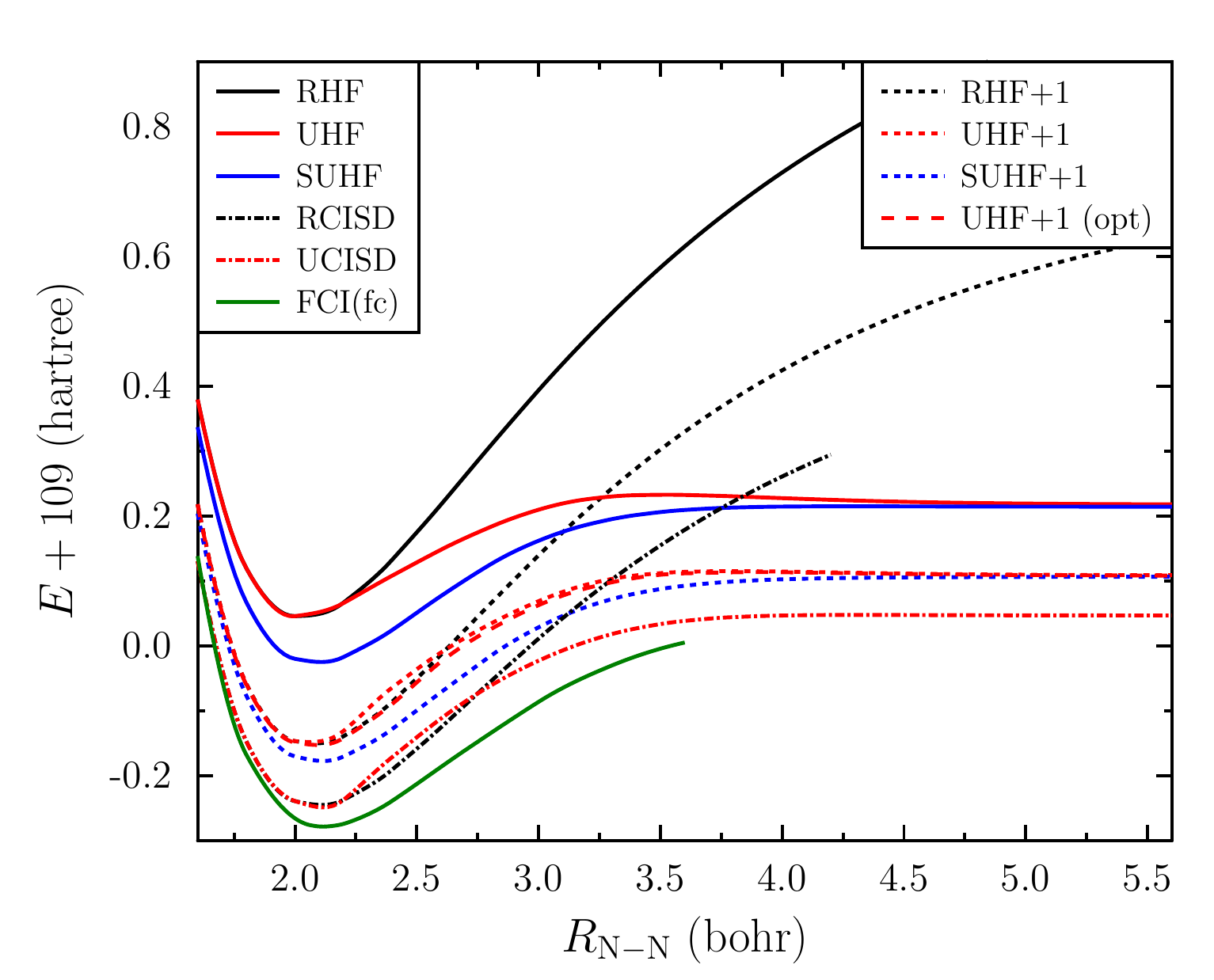}
  \caption{Dissociation profile of the N$_2$ molecule evaluated with
    various methods and the cc-pVDZ basis set. RHF+1 indicates the
    energy after a single gd step starting from the RHF
    wavefunction. UHF+1 (opt) indicates that the functional $E_1$ of
    Eq. \ref{eq:e1} is minimized using a trial wavefunction of UHF
    form. The FCI results, using the frozen-core approximation, are
    from Ref. \onlinecite{olsen2001}. \label{fig:n2}}
\end{figure}

We now consider the dissociation profile of N$_2$ with a cc-pVDZ basis
set. Reference FCI (with a frozen-core approximation) results for this
system are available from Ref. \onlinecite{olsen2001}. In
Fig. \ref{fig:n2}, we show the profiles obtained with RHF, UHF, and
SUHF, as well as the corresponding profiles after a single gd step is
performed (RHF+1, UHF+1, SUHF+1). Complementary to the RHF and UHF
results we show configuration interaction singles and doubles (CISD)
results starting from RHF and UHF, calculated using Gaussian 16
\cite{g16}. We note that it can be shown that the RHF+1 and UHF+1
energies should be above the corresponding RCISD and UCISD ones, as
the latter involve optimization of the amplitudes in the orthogonal
complement.

We emphasize that, in this case, we have produced the energies after a
single gd step without a formal vector representation of the starting
reference wavefunctions (in a basis of Slater determinants, the
dimension of the FCI vector is $\approx 1.4 \times 10^{12}$). In
evaluating the energy after the single gd step we computed $\langle 0
| H^3 |0 \rangle$ for a reference wavefunction that is either a single
Slater determinant or a linear combination of non-orthogonal
determinants. In both cases the evaluation can be completed, by making
use of Wick's theorem, with a computational effort of $\mathcal{O}(N^3
M^3)$ (with $N$ being the number of electrons and $M$ the number of
virtual orbitals). For this system we do not show results after a
second step as we do not currently have code that can evaluate up to
$\langle 0 | H^5 |0 \rangle$.

In the case of RHF, a single gd step recovers $\approx 50~\%$ of the
correlation energy near equilibrium. Starting from the SUHF reference
wavefunction we again recover $\approx 50~\%$ of the missing energy
after a single gd step. As expected, RCISD and UCISD energies are
significantly lower than RHF+1 and UHF+1, although the curves are
quite parallel to each other.

As we have discussed above, $E_1$ (Eq. \ref{eq:e1}) is explicitly a
functional of $|0 \rangle$. We can therefore choose to optimize the
parameters in $|0 \rangle$ to minimize $E_1$, instead of $E_0$. We
have done so with a UHF functional form, as shown in Fig. \ref{fig:n2}
under the label UHF+1 (opt). We note that, in setting up the
optimization process, we evaluate the derivative of $E_1$
analytically, as explained in App. \ref{sec:e1_grad}. The gradient
evaluation can still be carried out with a computational effort of
$\mathcal{O}(N^3 M^3)$, {\em i.e.}, the same scaling as for the
energy. While in this case we only see a modest improvement associated
with the minimization of $E_1$ directly, it can still be a valuable
approach as this can simplify the evaluation of molecular properties
expressed as energy derivatives.

We show in Tab. \ref{tab} the spectroscopic constants of N$_2$, O$_2$,
and F$_2$, evaluated using the cc-pCVTZ basis set, from HF and SUHF
calculations with and without the single gd step. In the case of
N$_2$, the length of the FCI vector expressed in a basis of Slater
determinants is $\approx 2.9 \times 10^{19}$, far exceeding what is
currently feasible with any current deterministic FCI code. We do not
show results starting from a UHF reference wavefunction for O$_2$ (as
UHF does not dissociate correctly) or F$_2$ (as the molecule is
unbound with UHF). The SUHF calculations on O$_2$ use a reference
determinant with $m_s = 0$ (while still projecting to a triplet state)
in order to reach the best possible dissociation limit.

The calculations shown in Tab. \ref{tab} are compared to high-quality
internally contracted multi-reference CI (IC-MRCI) results from
Ref. \onlinecite{dunning1997}, as well as the experimental values
compiled in the same reference. In all cases the single gd step yields
improved spectroscopic constants over the results obtained with HF or
SUHF. While the amount of the missing correlation recovered is modest
in all cases (around $30\%$), it is still noteworthy given that a
single gd step was performed. A better reference wavefunction or a
further gd or qn step can significantly improve the results presented
in Tab. \ref{tab}.

\begin{table*}[!htb]
  \caption{Spectroscopic constants ($E_e$ in hartree, $D_e$ in
    kcal/mol, $r_e$ in \AA, and $\omega_e$ in cm$^{-1}$) of N$_2$,
    O$_2$, and F$_2$ computed with various methods using the cc-pCVTZ
    basis set.  \label{tab}}
  \begin{ruledtabular}
    \begin{tabular}{l r r r r r r r r r r r r}
    & \multicolumn{4}{r}{N$_2$}
    & \multicolumn{4}{r}{O$_2$}
    & \multicolumn{4}{r}{F$_2$} \\\cline{2-5} \cline{6-9} \cline{10-13}
    method &
    $E_e$ &
    $D_e$ &
    $r_e$ &
    $\omega_e$ &
    $E_e$ &
    $D_e$ &
    $r_e$ &
    $\omega_e$ &
    $E_e$ &
    $D_e$ &
    $r_e$ &
    $\omega_e$ \\[2pt] \hline
    HF      & -108.987698 & 116.7 & 1.0660 & 2727.7 \\
    HF+1    & -109.081335 & 124.2 & 1.0642 & 2719.9 \\
    SUHF    & -109.057362 & 158.2 & 1.0909 & 2417.8 & -149.675867 &  30.8 & 1.1566 & 1980.4 & -198.851568 & 14.1 & 1.4803 & 663.9 \\
    SUHF+1  & -109.146989 & 163.2 & 1.0880 & 2435.5 & -149.777920 &  36.0 & 1.1541 & 1971.8 & -198.961698 & 15.1 & 1.4709 & 676.6 \\
    IC-MRCI\footnote{All-electron internally contracted multi-reference configuration interaction results from Ref. \onlinecite{dunning1997}.}   & -109.464001 & 220.0 & 1.1007 & 2350.7 & -150.209357 & 112.6 & 1.2099 & 1575.0 & -199.376661 & 32.3 & 1.4165 & 893.3 \\
    exptl\footnote{Experimental results, as compiled in Ref. \onlinecite{dunning1997}.} & & 228.4 & 1.0977 & 2358.6 & & 120.6 & 1.2075 & 1580.2 & & 39.0 & 1.4119 & 916.6
    \end{tabular}
  \end{ruledtabular}
\end{table*}

\section{Conclusions}
\label{sec:conclusions}

In this work, we have considered gradient descent and quasi-Newton
algorithms to reach the FCI ground state energy starting from an
arbitrary reference wavefunction. The FCI ground state is
systematically approached with increasing number of steps taken using
either algorithm.

A central ingredient of this work is to avoid an explicit
representation of the wavefunction, opting to write the energies, at
each step in the optimization, as functionals of the reference
wavefunction.  The resulting functionals have some useful properties:
\begin{itemize}
  \item They are independent of the form of the reference
    wavefunction. Therefore, they provide an unbiased way to compare
    how different wavefunctions evolve towards the FCI ground state
    along the optimization path.
  \item The functional forms are independent of the specific
    wavefunction parameters. That is to say, if the reference
    wavefunction is chosen as a single Slater determinant, then {\em
      any} Slater determinant (as long as it is not orthogonal to the
    FCI ground state) can be used with the functional forms provided.
\end{itemize}
While in this work we have focused on single determinant wavefunctions
as well as linear combinations of non-orthogonal determinants, the
evaluation of the required $\langle 0 |H^k |0 \rangle$ matrix elements
can also be efficiently done (in polynomial time) for other types of
wavefunctions such as multi-configuration self-consistent field
(MC-SCF) or generalized valence-bond (GVB) expansions, to name a
few. An efficient evaluation of the $\langle 0 |H^k |0 \rangle$ matrix
elements avoids the storage of intermediate wavefunctions as explicit
vectors over the Hilbert space, thus allowing the computation of
energies along the optimization path for systems where the dimension
of the Hilbert space is larger than the storage resources available.

Given that the evaluation of the matrix elements $\langle 0 | H^k | 0
\rangle$ can become quite expensive as $k$ increases, we realize that
practical applications may be limited to the first few steps. In this
case, a better starting initial inverse Hessian (as opposed to $B=I$)
can lead to substantially improved results. We are currently
investigating this possibility.

\section*{Data Availability}

The data that support the findings of this study are available from
the corresponding author upon reasonable request.

\begin{acknowledgments}
This work was supported by a generous start-up package from Wesleyan
University.
\end{acknowledgments}

\appendix

\section{Search direction from BFGS update}
\label{sec:bfgs}

We discuss in this appendix the form of the search direction $p_1 =
-B_1 g_1$, with $B_1$ constructed from from a
Broyden-Fletcher-Goldfarb-Shanno (BFGS)
\cite{broyden,fletcher,goldfarb,shanno} update formula (starting from
$B_0 = I$).

Let
\begin{align}
  s_0 &\equiv \, Z_1 - Z_0 = Z_1, \\
  y_0 &\equiv \, g_1 - g_0, 
\end{align}
which would yield $s_0= -\sigma_\ast g_0$ and $(y_0)_x =
(\alpha_1-\alpha_0) H_x^0 + \beta_1 H_x^y H_y^0$. With $\rho_0 =
1/[(s_0)^x (y_0)_x]$, the BFGS update takes the form
\begin{align}
  (B_1)_s^t &= \, (B_0)_s^t - \rho_0 (B_0)_s^x (y_0)_x (s_0)^t -
  \rho_0 (s_0)_s (y_0)^x (B_0)_x^t \nonumber \\
  &+ \, \rho_0^2 \Big[ \rho^{-1}_0 + (y_0)^x (B_0)_x^y (y_0)_y \Big]
  (s_0)_s (s_0)^t
\end{align}
We now carry an explicit evaluation of $p_1 = -B_1 g_1$. We note that
\begin{align}
  \rho^{-1}_0 &= \, -\sigma_\ast \alpha_0 (\alpha_1-\alpha_0) f_2
  -\sigma_\ast \alpha_0 \beta_1 f_3, \nonumber\\
  [y_0y_0] &\equiv \, (y_0)^x (y_0)_x \nonumber \\
  &=\, (\alpha_1-\alpha_0)^2 f_2 + 2(\alpha_1-\alpha_0)\beta_1 f_3 +
  \beta_1^2 f_4, \nonumber \\
  [s_0g_1] &\equiv \, (s_0)^x (g_1)_x \nonumber \\
  &= \, -\sigma_\ast \alpha_0 \alpha_1 f_2 - \sigma_\ast \alpha_0 \beta_1 f_3,
  \nonumber \\
  [y_0g_1] &\equiv \, (y_0)^x (g_1)_x \nonumber \\
  &= \, \alpha_1 (\alpha_1-\alpha_0) f_2
  -\alpha_0 \beta_1 f_3 + \beta_1^2 f_4. \nonumber
\end{align}
Therefore, $p_1$ takes the form
\begin{equation}
  (p_1)_x = \alpha'_1 H_x^0 + \beta'_1 H_x^y H_y^0,
\end{equation}
with
\begin{align}
  \alpha'_1 &=\, \alpha_1 - 2\sigma_\ast \rho_0^2 (\rho_0^{-1} + [y_0y_0])
         [s_0g_1] \nonumber \\
  &- \, \rho_0 (\alpha_1 - \alpha_0) [s_0g_1] + 2\sigma_\ast \rho_0
         [y_0g_1], \\ \beta'_1 &=\, \beta_1 - \rho_0 \beta_1 [s_0g_1].
\end{align}

\section{Optimization of $E_1$ with a Slater determinant}
\label{sec:e1_grad}

In this section we consider the optimization of the $E_1$ functional
of Eq. \ref{eq:e1} using a Slater determinant trial wavefunction. A
Slater determinant wavefunction $|\Phi \rangle$ can be parametrized in
terms of Thouless rotations (see, {\em e.g.},
Ref. \onlinecite{jimenez2012b}) as
\begin{equation}
  |\Phi \rangle = \mathcal{N} \exp \left(\sum_i T_{ai} c^\dagger_a \,
  c_i\right) |\Phi_0 \rangle,
  \label{eq:thou}
\end{equation}
where $\mathcal{N}$ is a normalization factor, $|\Phi_0 \rangle$ is
some initial reference Slater determinant, and $T_{ai}$ are Thouless
amplitudes to be optimized. Here, indices $a$ and $i$ run over
particle and holes, respectively.

The functional $E_1$ is fully expressed in terms of the expectation
values $\langle 0| H |0 \rangle$, $\langle 0| H^2 |0 \rangle$, and
$\langle 0| H^3 |0 \rangle$. In assembling the gradient of $E_1$, it
is sufficient to consider the gradients of those matrix
elements. Consider the matrix element $\langle 0|H^k |0 \rangle$, with
$|0\rangle = |\Phi \rangle$ and $|\Phi \rangle$ written in the form of
Eq. \ref{eq:thou}. The matrix element is given by
\begin{equation}
  \frac{\langle \Phi | H^k | \Phi \rangle}{\langle \Phi | \Phi \rangle}
  \equiv \frac{\langle \Phi_0 | \exp(T^\dagger) \, H^k \exp(T) | \Phi_0
    \rangle}{\langle \Phi_0 | \exp(T^\dagger) \, \exp(T)| \Phi_0
    \rangle}.
\end{equation}
Explicit differentiation of the equation above yields
\begin{equation}
  \left. \frac{\partial }{\partial T_{ai}^\ast} \frac{\langle \Phi | H^k
    | \Phi \rangle}{\langle \Phi | \Phi \rangle} \right|_{T = 0} =
  \frac{\langle \Phi_0 | c^\dagger_i \, c_a H^k | \Phi_0
    \rangle}{\langle \Phi_0 | \Phi_0 \rangle}.
\end{equation}

We finally note that the evaluation of $\langle \Phi |H^k |\Phi
\rangle$ and its derivatives, with $|0 \rangle$ chosen as a Slater
determinant, can be carried out by use of Wick's theorem.


\end{document}